\documentclass[pre,aps,onecolumn,showpacs,floatfix,superscriptaddress]{revtex4-1}
\usepackage{graphicx}
\usepackage{dcolumn}
\usepackage{color}
\usepackage{hyperref}
\usepackage{latexsym}
\usepackage{amsmath, amsthm, amssymb}
\usepackage{epsfig}
\usepackage{latexsym}
\usepackage{bm}
\def\be{\begin{equation}}
\def\ee{\end{equation}}
\begin{document}
\begin{center}
{\Large{\bf Directed Percolation and Directed Animals}} \\[2cm]
{\large{\bf Deepak Dhar}} \\
Theoretical Physics Group \\
Tata Institute of Fundamental Research \\
Homi Bhabha Road, Mumbai 400 005, ({\bf INDIA})\\[2cm]
\underbar{Abstract}
\end{center}

These lectures provide an introduction to the directed percolation and
directed animals problems, from a physicist's point of view.  The
probabilistic cellular automaton formulation of directed
percolation is introduced. The
planar duality of the diode-resistor-insulator percolation problem in
two dimensions, and relation of the directed percolation to undirected
first passage percolation problem are described.  Equivalence of the
$d$-dimensional directed animals problem to $(d-1)$-dimensional
Yang-Lee edge-singularity problem is established.  Self-organized
critical formulation of the percolation problem, which does not
involve any fine-tuning of coupling constants to get critical behavior
is briefly discussed.

\section{Directed Percolation}

Although directed percolation $(DP)$ was defined by Broadbent and
Hammersley along with undirected percolation in their first paper in
1957 [1], it did not receive much attention in percolation theory
until late seventies, when Blease obtained rather precise estimates of
some critical exponents [2] using the fact that one can efficiently
generate very long series expansions for this problem on the computer. 
Since then, it has been studied extensively, both as a simple model of
stochastic processes, and because of its applications in different
physical situations as diverse as star formation in galaxies [3],
reaction-diffusion systems [4], conduction in strong
electric field in semiconductors [5], and biological evolution [6]. 
The large number of papers devoted to $DP$ in current physics
literature are due to the fact that the critical behavior of a very
large number of stochastic evolving systems are in the DP universality 
class.  In this lecture, I briefly sketch
some of these connections.  Additional references may be found in
[7-9], or the book by Stauffer and Aharony [10]. 

\subsection{The Forest-Fire Model}

Can we embed the conventional bond percolation, say on a square
lattice,  in a stochastic process, which evolves randomly in time?
This may be done as follows: we think of the lattice as a forest, with
each site occupied by a tree.  A tree can be in one of three states:
green, burning or ash.  We assume that time $t=0$, there is a single
burning tree at the origin, and all other trees are green.  The time
evolution is discrete, and governed by the following rules:

\begin{enumerate}
\item[{(i)}] A burning tree at time $t$, becomes ash at time $(t+1)$.
\item[{(ii)}] A green tree not neighbouring any burning neighbours
remains green.  If it has $r$ burning neighbours at time $t$, it
catches fire and becomes a burning tree with probability $(1-q^r)$,
with $q = 1 - p$.
\item[{(iii)}] Ash remains ash for all subsequent times.
\end{enumerate}

\noindent Here $p$ is a parameter $0 < p < 1$.  If $p$ is small,
eventually the fire dies, and a configuration of only green trees and
ash results which does not change further.  It is easy to that the
probability that the final configuration of burnt trees is a given set
$S$, is the same as the probability that the cluster connected to the
origin is $S$ in the bond percolation problem.  There is a non zero
probability that fire survives infinitely long for $p > 1/2$, and this
exactly equals the probability of the origin belonging to the infinite
cluster. 

All questions relating to statistics of percolation clusters can be
reformulated as questions about the probabilities of different sink
states of the model.  The advantage of this formulation is that it
suggests other questions related to the time evolution of the system,
e.g. what is the velocity with which the fire front spreads outwards
for $p > p_c$? 

\subsection{Forest-Fire Model with Regrowth}

A somewhat different model results if we allow the burnt trees to
regrow, and become green again after some time-lag.  The main
qualitative difference over the previous case is that now the   
configuration with all sites green is the unique sink-state of the
system.  It
turns out that the precise duration of the time-lag is not very
important,
and it is just as well if we assume that burnt sites become green at
the next time step.  This allows us to work with only two states per
site: burning and green.  We replace the rule (i) in the previous model
by 
\begin{enumerate}
\item[{(i$'$)}] the burning site becomes green at the next time step 
\end{enumerate}
and we do not need rule (iii) any more.  Alternatively, this may be
thought of as a model of infection of disease in a population, where the 
infected individual usually recovers after a period of illness.    This is
the simplest formulation 
of the DP problem. It is interesting that this new model has a non-trivial
phase transition even in 1-dimension. It has been studied a
lot, and very precise estimates are available for critical parameters
and critical exponents [11], but an exact solution has not been
possible so far. 

\subsection{The Probabilistic Cellular Automaton Model}

Another terminology which is useful for the previous model is that of
a probabilistic cellular automaton [12].  At each site $i$, we have
discrete variable $n(i,t)$ at time $t$, taking values 0 or 1, ($0 =$
no fire, $1 =$ fire).  Evolution is discrete time, parallel and local. 
In the simplest case, $n(i,t+1)$ depends only on the value of
$[n(i-1,t) + n(i+1,t)]$.  Let $p_r$ be the probability that $n(i,t+1)$
is 1 if $n(i-1,t) + n(i+1,t) = r$.  We assume $p_0 = 0$.  This model
with two parameter $(p_1,p_2)$ is known  as the
Domany-Kinzel model [13].  For $p_1 = p_2 = p$, this model is the
directed site- percolation problem, and for $p_1 = p$, $p_2 = 2p -
p^2$ it corresponds to the directed bond- percolation at concentration
$p$.  Domany and Kinzel showed that this model can be solved exactly
for $p_2 = 1$, when the problem reduces to that of annihilating random
walkers on a line. 

\subsection{Relation to Quantum Spin Chains}

The directed percolation process is basically a Markov process that
involves allowing for the possibility of growth, propagation or death
of some activity (here fire).  In some applications, it is useful to think
of this in continous time.  Then, we say $n(i,t) = 1$, it can change
to 0 with rate 1, if $n(i,t) = 0$, and $n(i-1,t) + n(i+1,t) = r$ then
it can become 1 with rate $r\lambda$.

The
master equation for the evolution of ${\rm Prob}(C,t)$, the
probability that configuration of the system is $C$ at time $t$, which
has the form 
\be
{d \over dt} {\rm Prob}(C,t) = \sum_{C'} W_{CC'} {\rm Prob}(C',t),
\ee
can be thought of as a Schrodinger equation for the evolution of the
system.  The `wavefunction' to be in configuration $C$ at time $t$ is
${\rm Prob}(C,t)$, and the Hamiltonian is the matrix $W_{CC'}$.

One can then use the experience and insight gained from the study of
quantum Hamiltonians such as of spin-chains to learn about stochastic
evolving systems.  This has been a very useful approach in recent
years (see [14] for a  review), though this approach has a long
history (see, for example [15]).

In this particular case, if we think of $n_i = 1$ as spin up, and $n_i
= 0$ as spin down, then $W$ becomes the Hamiltonian of a linear chain
of spin-1/2 particles, with nearest neighbour couplings.  Introducing
$a^+_i$ and $a_i$ as the Pauli spin raising and lowering operators at
site $i$, we see that the Hamiltonian $W$ can be written in the form
\be
W = \sum_i \left[a_i + \lambda a^+_i a_i\left(  a^+_{i+1} +
 a^+_{i-1} \right) - a^+_i a_i \left(1 + \lambda a_{i+1}
a^+_{i+1} + \lambda a_{i-1} a^+_{i-1}\right)\right].
\ee
Note that $W$ is not hermitian.  This quantum mechanical formalism can
be developed further into quantum field theoretical formulation.  This
was historically the first technique used to estimate critical
exponents in all dimensions using the $\epsilon$-expansion techniques
(for references, see [9]).  We shall not discuss these further here.

\subsection{Scaling Theory for Directed Percolation}

The general theoretical treatment of the undirected percolation
(existence of critical threshold, exponential decay below criticality
etc.) goes through unchanged for the directed percolation problem.
There are two major differences: there is no unique infinite
percolation cluster.  The infinite cluster depends on the choice of
the origin.  Secondly, below but near the critical threshold $p_c$,
the clusters are large, but anisotropic.  We have to define two
different correlation lengths $\xi_\parallel$ and $\xi_\perp$, which
determine the average size of cluster along the preferred direction,
and transverse to it.  Near $p_c$, these lengths diverge as $(p_c -
p)^{-\nu_\parallel}$ and $(p_c - p)^{-\nu_\perp}$, where
$\nu_\parallel$ and $\nu_\perp$ are different exponents with
$\nu_\parallel > \nu_\perp$.

The probability that the site $(R_\perp, R_\parallel)$ is connected to
the origin, when the concentration is $p = p_c + \epsilon$ defines the
green's function $G(R_\perp, R_\parallel,\epsilon)$.  For small
$\epsilon$ and large $R_\perp$ and  $R_\parallel$ , this function is
expected to reduce to a scaling function of two
arguments
\be
G(R_\perp,R_\parallel,\epsilon) \approx
R_\parallel^{-\beta/\nu_\parallel} g(\epsilon
R_\perp^{1/\nu_\perp},\epsilon R_\parallel^{1/\nu_\parallel})
\ee
The function $g(x_1,x_2)$ is expected to decrease as
$\exp(-|x_2|^{\nu_\parallel})$
for $x_2 \rightarrow -\infty$, and increase as $x^\beta_2$ for $x
\rightarrow +\infty$.  For a fixed $\epsilon > 0$, the percolation
occurs within a cone of angle $\theta(p)$, and this angle varies as
$\epsilon^{\nu_\parallel - \nu_\perp}$ for small $\epsilon>0$.
Other
exponents such as $\gamma$ which characterizes  the divergence of  mean
cluster size near $p_c$ can be expressed in
terms of these three exponents $\beta, \nu_\perp$ and $\nu_\parallel$
by using scaling relations.

\subsection{Duality Transformation for Diode-Resistor-Insulator
Percolation in two dimensions}

One basic theoretical tool in the study of two-dimensional undirected
percolation the planar self-duality of the percolation problem.
Happily, this is generalized to the directed case.

The duality is best described in terms of a more general
diode-resistor-insulator percolation (with a suggestive acronym DRIP)
model[16].  We consider a square lattice.  Each bond is assumed to
independently occupied by a insulator, forward-biassed diode, reversed
biassed diode, or two-way conductor (resistor) with probabilities
$p_0, p_+, p_-$ and $p_2$ respectively $(p_0+ p_+ + p_- + p_2 = 1)$.
A
forward-biassed diode allows electric current to flow only up or
right, and a reversed-biassed diode only down or left.  The special
case $p_+ = p_- = 0$ corresponds to the usual undirected percolation,
and $p_2 = p_- = 0$ to the standard directed bond percolation. 

The duality transformation is a generalization of that for the
undirected problem.  To each insulating (resistor) bond, we dual bond
is resistor (insulating).  The dual of a left, right-, up- or
down-conducting doide bond is a diode conducting in the up-, down-,
right-
and left-direction respectively.  Clearly the 
resistor insulator percolation (RIP) to self-dual. It is easy to see that 
if there is an infinite directed path in the original model in some 
direction ( say, going towards up and right), then there is a blocking 
path in the dual problem along the same direction, which does not allow 
any connection across it ( in one direction). Thus spanning probabilities
in the original and dual problems get related. Similarly,  
the conductance of
a configuration can be similarly related to the conductance of the
dual lattice [16].

The dual of the diode-insulator percolation (DIP) problem $(p_+=p,
p_0=1-p, p_-= p_2= 0)$ is the diode-resistor percolation (DRP)
problem $(p_+ = p, p_2= 1-p, p_-= p_0= 0)$.  In the DRP problem,
for $p=1$ if we take a typical configuration, and consider the set of
sites reachable from the origin is just the first quadrant.  For
larger $p$, the cluster of wetted sites form a staircase with no
holes.  However for $p < p_c$, all sites of the lattice are wetted
with probability 1 in the thermodynamic limit.  The wedge angle of the
cluster of wetted sites increases from $\pi/2$ to $\pi$, as $p_2$
increases from 0 to $p_c$, and jumps to value $2\pi$ for all $p >
p_c$. Thus, the shape of the wetted cluster is apparently much simpler
in the DRP
problem than in its dual DIP problem.

\subsection{Relation between Directed and Undirected Problems}

The directed and undirected percolation problems belong to different
universality classes.  In addition, the directed percolation problem
shows extreme anisotropy near the critical point.  It is thus
interesting to realize that directed percolation properties are
implicitly present in the undirected percolation case, and one can
unravel these without 
any external imposition of a preferred direction.  It was
realized by Durrent and Liggett [17] that direction dependence of velocity
of wetted front in the undirected problem changes qualitively as the
threshold for directed percolation is crossed.  Consider undirected 
first-passage percolation on a square 
lattice. The time a fluid takes to wet a site after it has reached a 
neighbor is a random variable taking values 1 and 2 with probability $p$ 
and $(1-p)$ respectively. Then, if at time $t=0$, we start with a fluid 
source at origin, the size of wetted cluster upto time $t$ increases 
approximately linearly with $t$. Let $v(\theta)$ be the average velocity 
of the fluid front in the direction $\theta$. Then for $p$ above the 
directed percolation threshold $p_c$, $v(\theta)$ is exactly $1$ in any 
direction along which an infinite directed path exists. Behavior of this
velocity near $p_c$ can be described in terms of the standard DP exponents. 
For a more detailed discussion, see ref. [18].

\section{Directed animals and related models}

We start by defining the animal problem, and its relation to the
percolation problem in the more familiar undirected case first.

\subsection{Relation of Undirected Animals to the Percolation Problem}

   One of the basic objects of study  in  percolation
theory is $Prob(s,p)$, the probability that a randomly chosen site
belongs to a cluster having $s$ sites, where $p$ is  the concentration
of occupied sites (bonds). For $p < p_c$, there are few large clusters, and
for large $s$, ${\rm Prob}(s,p)$ varies as 
\be
{\rm Prob}(s,p < p_c) \sim  As^{-\theta} \exp[ - B(p)s].
\ee
At $p=p_c$, the behavior is a power-law
\be
{\rm Prob}(s,p_c) \sim A's^{-\tau}.
\ee
For $p >p_c$, there is an infinite cluster, but the number of large finite
clusters is again small. In this case, ${\rm Prob}(s,p)$ decreases as
a stretched exponential
\be
{\rm Prob}(s,p > p_c) \sim A''s^{-\theta'} exp[-B'(p)s^{\frac
{d-1}{d}}].
\ee

Here the functions $B(p)$ and $B'(p)$ depend on the details of the lattice
structure. These should go to zero as some universal power of
$(p-p_c)$ near the
critical percolation threshold. The exponents $\theta$ and $\theta'$ are
universal and are independent of $p$. 

While large finite clusters are not very likely for any $p$, we can still
ask what is the typical diameter of a cluster, given that it has $s$ sites
and has been drawn at random from a percolation problem at concentration
$p$. Consider, first, undirected percolation. For $p < p_c$, the
typical diameter $R_s$ varies as $K(p).s^{\nu}$ as
$s \rightarrow \infty$, where $K(p)$ is a p-dependent coefficient, but
the exponent $\nu$ is independent of $p$. At $p = p_c$, $R_s$ varies as a
different power $\nu_p$ of $s$. [Here $1/\nu_p$ is the fractal dimension
of percolation clusters.]  For $p > p_c$, $R$ varies as $s^{1/d}$.
Thus the structure of percolation clusters above $p_c$ is simple. The
structure of these clusters at $p_c$ is the subject of much study. The
animal problem deals
with the question of specifying structure of typical large clusters when
$p$ is below $p_c$. The simplest question is the value of the presumably
universal exponents $\theta$ and $\nu$.
 
As these exponent 
are independent of $p$, without loss of generality we may study it in the
special case $p \rightarrow 0$. In this limit, ${\rm Prob}(s,p)$ tends
to zero as $p^s$, but we get the simplification that all clusters of
$s$ sites occur with equal probability. This is known as the animal problem. 

Let $A_n$ be the number of distinct clusters of $n$ sites. For large $n$
this is expected to vary as
\be
A_n \sim  K \lambda^n n^{-\theta}
\ee
where $\lambda$ is a constant which depends on the lattice, and $\theta$
is an exponent. Giving equal weight to all the different animals of $n$
sites, we can determine the average size (say as measured by radius of
gyration)  of such animals. Call this
$R_n$. For large $n$,  $R_n$ varies as  $n^{\nu}$, where $\nu$ is an
exponent. Clearly, in $d$ dimensions $1 \leq \nu \leq 1/d$. The exact values
of $A_n$, or the value of $\lambda$ are  known only in one dimensions,  on
the Bethe lattice, and on some self-similar fractal lattices.  Parisi and
Sourlas \cite{parisi} have given heuristic arguments for remarkable
(presumably
exact)
relation between the exponents $\theta$ and $\nu$ 
\be
\theta = (d-2)\nu, ~~~~for ~~~   d \leq 8.
\ee
This is an analogue of the hyperscaling relation in the usual critical
phenomena, which relates the quantity $d\nu$ to thermodynamic exponents.
Note that here the factor which multiplies $\nu$ in this formula is
$(d-2)$, and not $d$. Thus the effective dimension of the system seems to
decrease from $d$ to $(d-2)$. This `dimensional reduction' is due to a
hidden supersymmetry in the field-theoretical formulation.
A rigorous justification for this argument is not
available yet. 

For 
$d=1,2,3$ and $4$, the exponent $\theta$ is believed to take the values
$-1,0,1/2$ and $5/6$. For $d \geq 8$, $\theta$ takes mean field value
$3/2$ and $\nu$ is $1/4$.

\subsection{Directed Animals}
  
As the undirected animals problem has not been solved exactly even in two
dimensions, it seems desirable to look at some simpler variants of the
problem. One simplification consists of making the bonds directed, i.e.
look for the statistics of directed percolation clusters. It turns out
that this variant is much more tractable analytically. In the following,
I briefly review known results on this problem.

Consider a $d$-dimensional hypercubical lattice. We assume $d=2$ for
simplicity. Each site $(x,y)$ has two bonds directed outwards towards the
sites $(x+1,y)$ and $(x,y+1)$. A directed (site-)animal is a set of
`occupied' sites
(including the origin) such that for each occupied site $(x,y)$, other
than the origin, at least one of the two sites $(x-1,y)$ and $(x,y-1)$ is
also occupied.   We shall denote the number of distinct animals having $s$ 
sites by $A_s$. For example, it is easy to verify, or write  a short
computer program for exhaustive enumeration of such animals 
\cite{rednerprogram}, and see that for $n=1,2,3,4,5 \ldots$ the numbers
$A_n$
are $1,2,5,13,35 \ldots$ . Based on the first few terms of this series,
Dhar et al \cite{dharPB1} were able to guess the exact formula
\be
    A_n = \int_0^{2\pi}  d\theta (1+cos\theta)(1+2 cos\theta)^{n-1}
\ee
The generating function of these animal numbers $A(x)=\sum_n {A_n x^n}$
satisfies a simple quadratic equation
\be
 (1-3x)[A(x)+A^2(x)] = x
\ee
We now indicate how this result comes about \cite{dhar4}. We note that in
a dirrected
animal, the allowed configuration of occupied sites on the line $x+y=T$
depend only on the configuration of occupied sites in the animal on the
line $x+y=T-1$. Starting with a single occupied site on the line $x+y=0$,
on the line $x+y=1$, we can have at most two occupied sites: $(1,0)$ and
$(0,1)$. This leads to the recursion equation
\be
A(x) = x[1 + 2 A(x) + A_{11}(x)]
\ee
where $A_{11}(x)$ is the generating function of animals starting from two
occupied sites $(1,0)$ and $(0,1)$. In general, if we define $A_C(x)$ as
the generating function of all animals starting with a source $C$ on the
line $x+y=T$, we get a recursion relation
\be
A_C(x) = x^{|C|} [ 1 + \sum_{C'}A_{C'}(x) ]
\ee

where $|C|$ is the number of occupied sites in $C$, and the sum over $C'$
is over all configurations of occupied sites on the line $x+y=T+1$
of a directed animal consistent with $C$. One can generate the animal
numbers $A_n$ for quite large values of $n$ ( $\sim 100$) using such
recursion relations in a computer program. Such series can then be
analysed using rather sophisticated extrapolation techniques. A large
number of such series are known by now (for some examples,
see \cite{conway}).

\subsection{Relation to Hard-Core lattice Gas models}

Alternatively, the above recursion relation may be viewed as the Chapman-
Kolmogorov equation for a probabilistic cellular automaton on a line
defined by the following rules: At time $\tau = 0$, all sites on 
the line $x+y=0$ are assumed to be empty. At (integer -valued) time $\tau$
sites along the line $x+y=-\tau$ are examined, and a site $(x,y)$ is
occupied with probability $p$ if both the sites $(x+1,y)$ and $(x,y+1)$
are empty. Else, the site is left empty. Clearly this corresponds to a
Domany-Kinzel type automaton with $p_0=p, p_1=p_2=0$. This may be thought
of as a model of growth of mixed crystals from solution layer by layer.
We start with a solution of, say,  NaCl and KCl. The crystal obtained by
evaporating such a solution is substitutionally disordered, with Na and K
atoms placed at random in the, in other ways regular, lattice-structure.
We ignore the chlorine atoms, and think of a simple cubic lattice formed 
by adding layers of Na and K atoms. We further assume that due to their
larger size, two K atoms cannot be adjacent to each other. If allowed to
be occupied by a K-atom, a position is actually occupied by a K atom with
probability $p$, else it is occupied by Na.  

Clearly, the probability that a site $P=(x,y)$ is occupied by a K-atom
equals $p$ multiplied the probability that both the sites $P'=(x+1,y)$ and
$P''=(x,y+1)$ are not occupied by K-atoms. Using the inclusion-exclusion
principle, we get
\be
Prob(A) = p [ 1 - Prob(A') - Prob(A'') + Prob(A'A'')]
\ee
which is of the same form as the recursion equation (11).
In general, for $Prob(C)$, the probability that all sites of the set $C$ 
lying on a line $x+y= constant$ are occupied, we get
\be
Prob(C) = p^{|C|} [ 1 + \sum_{C'} (-1)^{|C'|} Prob(C')]
\ee
where the sum over $C'$ is over all proper subsets of the set of backward
(in $\tau$) neighbors of $C$.  Comparing with eq.(12) we see that
\be
A_C(x=-p) = (-1)^{|C|} Prob(C)
\ee
With this relation, the problem of determing the animal generating
function $A(x)$ reduces to finding the average density of K-atoms in the
steady state of th probabilistic cellular automaton model defined above. 
The latter problem turns out to be very simple, as the corresponding rates
satisfy the detailed balance condition. The corresponding hamiltonian is
that of a  lattice gas on a linear lattice with nearest-neighbor
exclusion. The trivial exact solution of this gives us the exact result
(9), and we find that
\be
\theta = \nu_{\perp} = 1/2, ~~~~~~ for~~~~ d=2.
\ee
For details, consult \cite{dhar4}. Many more exact results for the
2-dimensional problem can be obtained in
this way. For a more recent paper, which contained some interesting
unproved
(so far) conjectures also, see \cite{conway}.
The mean longitudinal size of directed animals $R_{\|} \sim n^{\nu_{\|}}$.
The value of the exponent $\nu_{\|}$ is not known exactly. numerical
estimates suggest that it is not a simple fraction \cite{dhar5}.

The exact calculation of the partition function of the hard-hexagon
lattice gas by Baxter \cite{baxter} as a function of the activity can be
used to determine the exponents of the d=3 directed animals problem. This
gives
\be
\theta = 2 \nu_{\perp} = 5/6, ~~~~~~~~for ~~~~ d=3.
\ee

\subsection{Relation to the Lee-Yang Edge Singularity Problem}

We have seen that the generating function of animal numbers in
d-dimensions is becomes the expression for density of a nearest -neighbor
exclusion lattice gas in (d-1) dimensions expressed as a function of the
chemical activity. The latter is the well-known Mayer expansion. The
radius of convergence of this determined by the closest singularity to the
origin of the analytically continued free energy $f(z)$ as a function of
the activity $z$, now treated as a complex variable. Now, the
singularities of $f(z)$ come from the zeroes of the partition function.
It was noted by Lee and Yang long ago that these zeroes usually are
distributed continuously along some lines in the complex -$z$
plane. The line-density of zeroes on the line at the point $z$ near the
endpoint of a line $z_c$ varies
as a pwer of $(z-z_c)^\sigma$. This exponent $\sigma$ is quite universal,
and does not seem to depend on the details of the hamiltonian of the
system, only on the dimension of space $d$ \cite{laifisher}. The directed
animal exponent $\theta$  can be expressed in terms of this by the
relation
\be
\theta(d) = \sigma(d-1) + 1
\ee
Parisi and Sourlas related the $\theta_{undir}$ exponent
for the undirected animals problem in dimension
$d+2$ to the Lee-Yang Edge-singularity problem in $d$ dimensions
\be
\theta_{undir}(d+2) = \sigma(d) +2
\ee
Knowing the value of $\theta(d=3) = 5/6$, we see using these relations
that $\sigma(d=2)=-1/6$ and $\theta_{undir}(d=4)=11/6$.

\section{Self-organized Criticality}

We have seen that in the percolation problem, the correlations decay
as power-laws, and the correlation length is infinite {\it precisely
at the critical point}. At the critical point, the clusters have a
fractal structure. Away from the critical point, the correlation
function decays exponentially with distance. The same thing occurs in
many other lattice models in statistical physics with discrete degrees
of freedom per site with local interactions, like the Ising model. The
equilibrium state of these models corresponds to exponential decay of
correlations, {\it except at the critical point}.  Thus, in order to
get power-law correlations, one has to fine-tune the coupling constant
to be very near the critical value. Study of behavior of systems near
the critical point has been a major topic of study in statistical
physics in the last three decades, and a fair understanding has
emerged of the universality of, and relations between, critical
exponents from the renormalization-group approach to critical
phenomena pioneered by K.  Wilson. 

On the other hand, in nature, we find a large number of fractal
objects such as clouds, mountains, river-networks, which seem to be
characterized by power-law correlations over a fairly broad range of
length scales. But in these cases, it is clear that the physical
processes that give rise to such structures must not require any
fine-tuning of any control parameters. This realization has led to a
lot of interest in the physics community in the study of systems which
reach a staedy state with long-range correlations without any need for
the 'unnatural' fine-tuning of any control parameters. These systems
may be said to self-organize into a critical state, and have been
called Self-Organized Critical systems (SOC) \cite{bak}. 

In this section, I will try to introduce some percolation models
showing criticality without any finetuning of control parameters. 

\subsection{Invasion Percolation}

The simplest percolation model of this type is the
invasion-percolation model. This was introduced by Wilkinson and
Willemson in 1983 \cite{wilkinson}, and precedes the enunciation of
the general concepts of SOC by Bak et al. 

We consider a network of pipes forming a grid, say a two dimensional 
square lattice. The diameters of pipes are assumed to be independent, 
identically distributed random variables with some specified
continuous distribution. Now we imagine forcing a viscous fluid into
an initially empty network from outside, inserting it at one of the
nodes, say the origin. [In actual applications, one pushes steam in a
porous oil-bearing rock, forcing out oil.] With time, this fluid will
spread via the network of pipes. As it is harder to push fluid in a
thinner pipe, this spreading occurs preferentially using a subset of
thicker pipes. At any time $t$, there is a set of pipes $B_t$ which
lies at the boundary seperating the wet from the dry nodes. We assume
that actual spreading will occur using the thickest pipe from the set
$B_t$. We may characterize the amount of force needed to push fluid
through the pipe (overcoming the capillary forces) using a real
variable $x$ with $0 \leq x \leq 1$, such that lower value of $x$
implies lower force.  The time may be measured in discrete 
units in terms of the number of sites wetted upto that instant. 

Thus in this model, the cluster of wetted sites grows in time, always 
using the weakest of the available links for growth. However the shape
of the cluster of wetted sites is stochastic, and depends on the
realization of the random medium.  The wetted cluster can have holes
which may not be filled for a long time.  In fact, simulations show
that for large times $t$, the cluster $C_t$ of sites wetted up to time
$t$ contains holes of all length scales, and looks like incipient
infinite percolation cluster.  

The relation of this model to the percolation problem becomes
clearer,if we assume that probability distribution of the variable $x$
is uniform in the interval [0,1]. This we can do without loss of
generality, as the variable $x$ can be replaced by any monotonic
increasing function of $x$ without affecting the dynamics. Let the
percolation threshold on this lattice be $p_c$. Consider a value $p_1
= p_c + \epsilon $, where $\epsilon$ is an arbitrarily small positive
number, and consider the infinite percolation cluster formed using
only the pipes for which $x \leq p_1$. Once the cluster of wetted
sites hits any of the sites of this cluster (as it must eventually),
further growth can occur only using links of this infinite cluster.
Thus, if ${\rm Prob}(x)$ is the limiting probability distribution at large
times that the next growth occurs using a bond having value $x$, we
conclude that ${\rm Prob}(x)$ is exactly zero for all $x \geq p_c$. On the
other hand, ${\rm Prob}(x)$ must be nonzero for all $x \leq p_c$, as there
are no infinite clusters in which all bonds have a value $x < p_c
-\epsilon$.  

The interesting point about this model is that the dynamical rules
make no mention of $p_c$. Thus no unnatural fine-tuning of parameters is
done to get the fractal growth cluster, whose properties at large
length scales are same as of the critical percolation clusters. For
finite times $t$, the distribution of clusters $C_t$ is not same as in
the standard percolation problem, as is easy to check for
$t=1,2,\cdots$.  

\subsection{The Sneppen model}

We may similarly study the invasion percolation problem for the
diode-resistor network defined earlier. Consider,for simplicity, again
a square lattice. We assume that all links allow fuid to flow easily
(no force) in the positive direction (up or right), but need a finite
force $x$ to flow in the reverse direction. The values of $x$ are
i.i.d. random variables for each link, with a uniform distribution
between 0 and 1. We imagine that time $t=0$ all sites $(x,y)$ with
$x+y \geq 0$ are wet. At any time $t$, we choose the bond whith the
least value of $x$ lying on the boundary between the wet and dry
sites, and push fluid trough it. Thus the wetted region grows, and the
interface between wet and dry sites moves left and down. If a site is
wetted but its rightward or upward neighbor is dry, then these sites
also become wet.  

This model is usually called the Sneppen model \cite{sneppen}. The
important point is that the movement of the interface occurs in
bursts, and there is a distribution of burst sizes with a power-law
tail. If you look at the values of $x$ selected, they will lie only
between 0 and $p_c$, where $p_c$ is the threshold for the
diode-resistor percolation in this model. The surface after a long
time becomes 'rough', with a nontrivial roughness exponent, which is
relatable to exponents of directed percolation. There has been a lot
of interest in determining the distribution of burst sizes in this
model in 2 and higher dimensions in recent years \cite{buldyrev}.

\subsection{Self-organized Directed Percolation}

In the Sneppen model, one does not need fine-tuning to get a critical 
interface. The basic mechanism which makes this possible is the fact
that the growth is assumed to occur at the site corresponding to
global minimum of all $\{x\}$ along the interface. This feature, while
no doubt a good approximation in some situations, is not very
aesthetically pleasing, as it implies that all points at the interface
`know' about the status of the interface at all points.  In physical
systems, one usually prefers to write evolution laws which are local,
and depend only on values of various quantities in the neighbourhood
of the point. Can one make a model with local stochastic evolution
rules, which gives nontrivial critical behavior without any
fine-tuning of control parameters? 

It was realized by Grassberger and Zhang \cite{grassberger} that
this is indeed possible. They considered a simple coupled map lattice
defined as follows: Consider a linear chain with a real variable
$x(i,t)$ at each site $i$  with $ 0 \leq x(i,t) \leq 1$. At time
$t=0$, all the variables $x(i,t=0)=0$. Time is discrete, and all site
evolve in parallel using the rule 
\be 
x(i,t+1)= Max [ \eta(i,t), Min( x(i,t), x(i-1,t)] 
\ee
where $\eta(i,t)$ are random variables drawn from a uniform
distribution between $0$ and $1$, independent for different space-time
points $(i,t)$.  After a time $T$, all the values $x(i,T)$ have an
marginal distribution given by
\be
{\rm Prob}(x(i,T) \geq p) = P_{DP}(p,T)
\ee
where $P_{DP}(p,T)$ is the probability that site $i$ will be wet at
time $T$ for a directed site percolation problem on the square lattice
with site concentration $p$ when the initial state is all sites wet.
To see this, we just note that if we define a variable $y(i,t)$, which
takes values 1 and 0 according as $x(i,t)$ is $\leq p$ or greater than
$p$. Then the process $y(i,t)$ is a simple DP percolation process [1
corresponds to a wet site, 0 to healthy]. The rate of approach of the
distribution of $x(i,T)$ to the limiting distribution gives
information about other DP exponents. 

\subsection{Self-Organized Undirected Percolation}

Can one make a similar model for undirected percolation? This is also
possible but it involves a non-trivial variation of the
Grassberger-Zhang construction. Consider a network of sites and bonds,
where each bond is randomly assigned a strength $x$ lying between $0$
and $1$.  We assign a variable $y(i,t)$ to each site $i$, also lying
between $0$ and $1$ which evolves with time $t$ by the following
rules:  

\begin{enumerate}
\item[{i)}] For all sites $i$ not at the boundary, $y(i,t=0)=1$. At
boundary sites $y(i,t)=0$ for all times $t$. 

\item[{ii)}] All sites not at the boundary  are updated in parallel
using the rule: 
\be
y(i,t+1)=Min_j[Max(y(j,t),x_{ij})]
\ee
where the minimum is taken over all sites $j$ neighboring $i$, and
$x_{ij}$ is the strength of bond lying between $i$ and $j$.
\end{enumerate}

It is easily seen that $y$'s are nonincreasing functions of time, and
for any finite lattice, they reach some fixed point values $y^*(i)$,
which depend on the configuration. By induction on $t$, it is easy to
show that for all times $t$ and all sites $i$,
there exists a
path from $i$ to the boundary which uses only bonds with strength
$\leq y(i,t)$. As in the directed case, the limiting distribution
${\rm Prob}(y^* \geq z)$ of $y^*$'s is equal to the fraction of sites which
belong to the infinite cluster for bond concentration $p=z$.

More generally, can one take a usual equilibrium statistical model
with a critical point, say an Ising model, and endow it with local
stochastic evolution rules such that it will at large times always
relax to a steady state which corresponds to the model at its critical
temperature without any fine-tuning of parameters? These questions are
still open. This has been possible so far only in a model with a rather
complicated set of variables at each site \cite{bagnoli}. 
 
It is presumably clear from the examples discussed above that the
ideas of percolation theory have found many uses in many ways in
physics, and continue to do so. It is hoped that these will also
provide new directions to the more mathematical studies.

\end{document}